\documentclass[11pt,letterpaper]{article}
\pdfoutput=1

\usepackage{graphicx,array}
\usepackage[dvipsnames]{xcolor}
\definecolor{darkblue}{rgb}{0,0.1,0.5}
\definecolor{darkgreen}{rgb}{0,0.5,0.2}
\definecolor{darkred}{RGB}{153,26,0}
\usepackage{ulem}
\usepackage{slashed}
\definecolor{seablue}{rgb}{0,0.2,0.6}
\usepackage{latexsym}
\usepackage{amssymb, amsmath}
\usepackage{slashed}
\usepackage{bm}        % for math
\usepackage[numbers,sort&compress]{natbib}
\usepackage{bm,relsize}
\usepackage{slashed}
\definecolor{viola}{RGB}{134,41,198}
\usepackage{mathrsfs}
\usepackage{hyperref} %Automatically links \label and \ref commands; Always load last
\hypersetup{
    colorlinks=true,       % false: boxed links; true: colored links
    linkcolor=darkblue,          % color of internal links
    citecolor=BrickRed,        % color of links to bibliography
    filecolor=magenta,      % color of file links
    urlcolor=MidnightBlue           % color of external links
}
\usepackage[all]{hypcap} %Link navagates to top of figure instead of caption (below fig)
\usepackage{subcaption}

\usepackage{listings}
\usepackage{xcolor}

\definecolor{codebg}{rgb}{0.95,0.95,0.95}

\usepackage{colortbl}

\usepackage[font=small,labelfont=bf,labelsep=period]{caption}

\newcommand{\be}{\begin{equation}}
\newcommand{\ee}{\end{equation}}

\usepackage{tikzfeynman}

\begin{document}

%%%%%%%%%%%%%%%%%%%%%%%%%%%%%%%%%%%%%%%%%%%%%%%%%%%%%%%%%%%%%%%%%%%%%%%%%%
\begin{flushright}

\end{flushright}
\vspace{.6cm}
\begin{center}
{\LARGE \bf 
%Primordial Black Holes and Gravitational Waves\\
%in Slow Roll Single Field Inflation
Reconstructing the Inflaton Potential:\\
Primordial Black Holes and Gravitational Waves\\[3mm]
in Slow Roll and Ultra Slow Roll Single Field Inflation 
}\\
\bigskip\vspace{1cm}
{
\large Gabriele Autieri$^{a,b}$, Michele Redi$^c$}
\\[7mm]
 {\it \small
$^a$ SISSA International School for Advanced Studies, Via Bonomea 265, 34136, Trieste, Italy\\
\vspace{.1cm}
$^b$ INFN  Sezione di Trieste, Via Bonomea 265, 34136, Trieste, Italy\\
\vspace{.1cm}
$^c$ INFN Sezione di Firenze, Via G. Sansone 1, I-50019 Sesto Fiorentino, Italy
 }
\end{center}

\vspace{2cm}

\centerline{\bf Abstract} 
\begin{quote}
We present new single field inflationary scenarios that produce the critical abundance of primordial black holes as dark matter reconstructing the inflaton potential from an input power spectrum. The method is exact in the slow roll approximation but remains effective even when the slow roll conditions are temporarily violated such as in ultra slow roll models. With this method we construct new ultra slow roll scenarios and also models that reproduce the DM abundance within the slow roll regime. As a second application we consider a scalar power spectrum that generates a secondary gravitational wave background compatible with the one recently observed in Pulsar Timing Arrays experiments. These scenarios could be tested by future observations of $\mu-$distortions of the CMB.
\end{quote}

\vfill
\noindent\line(1,0){188}
{\scriptsize{ \\ E-mail:\texttt{  \href{mailto:michele.redi@fi.infn.it}{michele.redi@fi.infn.it}, \href{gautieri@sissa.it}{gautieri@sissa.it}}}}

\newpage

\tableofcontents

\section{Introduction}

The only plausible dark matter (DM) candidate within the Standard Model is represented by primordial black holes (PBH) produced in the early universe from the gravitational collapse of large density perturbations \cite{Carr:1974nx}. The abundance of  these objects is however negligible within the vanilla cosmological scenario $\Lambda$CDM so that new physics is necessary to explain the observed DM abundance as made of PBH. 

One of the most attractive production mechanisms of PBH is from the collapse of inflationary perturbations as they re-enter the horizon during standard cosmology, see \cite{Ozsoy:2023ryl,Khlopov:2008qy,Choudhury:2024aji} for reviews. As is well known inflation is extremely successful in explaining the initial conditions of our universe predicting in particular the seeds of perturbations that eventually grow under the pull of gravity to give rise to the complex structures that we observe today. Remarkably all the measurements can be explained in terms of an almost scale invariant primordial power spectrum with an amplitude of order $10^{-9}$ and small tilt. This is easily achieved in slow roll (SR) inflationary scenarios  where the inflaton slowly goes down a flat potential.
All the observations however only test the first few e-foldings of inflation, corresponding to cosmological distances in the present  universe. This leaves enormous room for new phenomena to occur during the remaining $O(50)$  e-foldings of inflation, corresponding  to highly non-linear scales in the present universe.

The observation above is instrumental for the production of PBHs and gravitational wave (GW) backgrounds as a secondary effect. The primordial scalar power spectrum could deviate from the almost scale invariant one measured at cosmological scales and be enhanced at short distances. If the curvature perturbations become larger than a certain threshold \cite{Nakama:2013ica,Musco:2018rwt}, when they re-enter the horizon they will produce PBHs of size similar to horizon at that time. The power spectrum can thus be mapped into the mass and abundance of PBHs in our universe.

Production of PBH requires an enhancement of the power spectrum of about 7 orders of magnitude compared to the one measured through the Cosmic Microwave Background (CMB). The comoving scale where the spectrum is  enhanced is directly linked to the mass of the PBH produced. From an observational point of view DM made of PBH requires $k_p\sim 10^{12-14}\,{\rm Mpc}^{-1}$ \cite{Green:2020jor}. The enhancement can be generated in single field inflationary models if the inflaton crosses an extremely flat region of the potential. Often in the literature one considers the ultra-slow roll (USR) regime \cite{Tsamis:2003px,Kinney:2005vj} obtained by setting the first derivative of the inflaton potential to zero where analytical solutions are also available. In this case the power spectrum grows as $k^4$ so that the amplitude to produce PBHs is obtained after a couple of e-foldings. This rapid growth violates slow conditions (SR) of the inflaton in the transition region. 
 
In this work we explore more general single field inflationary scenarios that deviate from USR models and produce an enhanced power spectrum, see \cite{Ballesteros:2017fsr,Franciolini:2022pav} for other realizations. To do so we start from a desired power spectrum and derive a potential that reproduces the enhancement. This can be done very simply in the slow roll approximation where the power spectrum can be directly mapped to a potential. We show in particular that the critical abundance of DM made of PBH can be obtained within the SR regime. This requires a milder growth  of the power spectrum compared to USR models that thus extends to larger scales and could be possibly observable. This question was also explored in \cite{Motohashi:2017kbs} with a different conclusion. 

If the enhancement takes place very rapidly the potential that we derive violates SR conditions as USR models.  In this case the true power spectrum is modified with respect to the input one but it still produces the required enhancement. This produces new USR models.

As a second application of our potential reconstruction we consider the production of a stochastic GW background. Recently Pulsar Timing Arrays experiments \cite{NG15-1,NG15-2,NG15-3,NG15-4,EPTA-1,EPTA-2,EPTA-3,PPTA-1,PPTA-2,PPTA-3,CPTA} have reported convincing evidence of a GW background with frequencies $f\sim 10^{-8}$ Hz and energy fraction $\Omega_{\rm GW}\sim  10^{-10}$. Tensor perturbations produced during inflation are too small to generate such a background. However an enhanced power spectrum of scalar perturbations can generate at second order such a background for a scalar amplitude $O(10^{-2})$. 
To reproduce the frequency the spectrum should be enhanced at  scales $k\sim 10^7\,{\rm Mpc}^{-1}$. We thus apply our method to construct inflationary potentials both in SR or USR regimes. In the first case the scalar power spectrum deviates from the scale invariant one at cosmological scales.  This leads to significant constraints from CMB, $\mu$-distortions \cite{Fixsen:1996nj} and possibly dynamical heating of stars \cite{Graham:2024hah}. The enhanced power spectrum could also produce GWs at higher frequencies, relevant for  experiments such as LISA \cite{LISA}.

The paper is organized as follows. In section \ref{sec:method} we explain how to construct a potential that produces an  input power spectrum in the SR regime. In section 4 we apply this method to the production of PBH from inflationary perturbations, showing that the critical abundance can be obtained both in slow  and USR regime. Section \ref{sec:PTA} is devoted to GW from secondary GWs produced from inflation. We present examples where the GW background is produced in SR and USR regime. We conclude in section 5 with outlook on future directions.

\section{Building an enhanced power spectrum}
\label{sec:method}

We start describing a simple algorithm to construct single field inflationary scenarios that produce an enhanced primordial power spectrum at a given scale. The method is very general and can be applied to design specific features during inflation.

We consider single field inflationary scenarios described by a scalar field,
\begin{equation}
{\cal L}=\frac 1 2 (\partial \phi)^2- V(\phi)
\end{equation}
Deviations from exact de-Sitter expansion are measured by $\epsilon \equiv -\dot H/H^2$. In the SR regime where $\epsilon<1$ the power spectrum is given by,
\begin{equation}
\mathcal{P}(k) = \frac {1}{2 \epsilon_{\rm SR} M_p^2} \left( \frac {H_I}{2\pi}\right)^2\Big|_{k/a=H_I}\,,~~~~~~~~~~\epsilon_{\rm SR}= \frac {M_p^2} 2 \left(\frac{V_\phi}V\right)^2
\label{eq:slowrollpower}
\end{equation}
where all quantities are evaluated at horizon crossing. In what follows we parametrize $k=k_* e^N$ with $k_*=0.05\, {\rm Mpc}^{-1}$ the CMB pivot scale, so that the mode $k$ crosses the horizon after $N$ e-foldings of visible inflation. Note that SR requires not only $\epsilon$ to be small but also higher order terms, in particular the second slow roll parameter,
\begin{equation}
\eta\equiv - \frac {\ddot{H}}{2 H \dot{H}}=\epsilon -\frac 1 2 \frac {d \log \epsilon}{dN}
    \label{eq:eta}
\end{equation}
Actually, as emphasized in \cite{Motohashi:2017kbs}, eq. (\ref{eq:slowrollpower}) holds for $\epsilon<1$ only if  the second SR condition, $|\eta|<1$, is also satisfied.

Independently of SR dynamics, for a given input power spectrum, we can define $\epsilon(N)$ from eq. (\ref{eq:slowrollpower}) assuming $H_I$ to be constant, see Ref. \cite{Hertzberg:2017dkh} for a related approach. Specifically we define,
\begin{equation}
\epsilon(N)\equiv \frac 1 {2 M_p^2} \left(\frac {d\phi}{dN}\right)^2 = \epsilon(N)_{\rm SR}
\label{eq:eps}
\end{equation}
The first equality is an exact equation valid for single field inflationary models and we require this to be equal to the function $\epsilon(N)_{\rm SR}$ obtained from eq. (\ref{eq:slowrollpower}). The $\eta$ parameter is also determined as,
\begin{equation}
\eta(N)=\epsilon(N) - \frac {dN}{d\phi} \frac{d^2\phi}{dN^2}  
\label{eq:etaexact}
\end{equation}

This produces a model where the exact $\epsilon$ is enhanced by the required amount\footnote{An alternative approach would be to derive the potential integrating $\epsilon_{\rm SR}= M_p^2/2 (V_\phi/V)^2$ (see \cite{Lidsey:1995np} for a similar approach). This would work in the SR regime but it fails to produce a model with the required enhancement if the growth of the power spectrum is fast.}. To determine the potential that produces $\epsilon$ one can proceed as in Ref. \cite{UF22}. The classical equation of motion for the inflaton can be written as,
\begin{equation}
\frac {d^2\phi}{dN^2}+(3-\epsilon)\frac {d\phi}{dN}=M_p^2(\epsilon-3)\frac {V_\phi}{V}
\end{equation}
Since $\phi(N)$ is a known function this equation  can be integrated to determine  $V(N)$ that together with eq. (\ref{eq:eps}) determines the potential $V(\phi)$. Explicitly,
\begin{equation}
\begin{split}
&V(N)=V(N_0) \exp\left[-2 \int_{N_0}^N dN'\frac{\epsilon(3-\eta)}{3-\epsilon}\right]\\
&\phi(N)= \phi_0+ M_p\int_{N_0}^N dN' \sqrt{2\epsilon}
\end{split}
\label{eq:potential reconstruction}
\end{equation}
that can be solved numerically. We emphasize that these equations are valid in general beyond SR dynamics.

The procedure above would be exact in the SR approximation. If the latter is violated, as for example in USR scenarios where $\eta\approx 3$ in the transition region, or only approximately satisfied, the actual power spectrum will be different from the input one. To address this issue, for the reconstructed potential (\ref{eq:potential reconstruction}) one can compute the power spectrum by solving the Mukhanov-Sasaki equation for the perturbations. Denoting the curvature perturbations $\zeta_k= \sqrt{2 \epsilon}a  M_p v_k$ they satisfy,
\begin{equation}
\label{eq:MS}
\begin{split}
&\frac {d^2v_k}{dN^2}+(1-\epsilon) \frac {dv_k}{dN}+ \left[\frac {k^2}{(a H_I)^2}+(1+\epsilon-\eta)(\eta-2) - \frac d{dN} (\epsilon-\eta) \right] v_k=0
\nonumber \\
&v_k(N_k)=\frac 1 {\sqrt{2k}}\,,~~~~~~~~~~~~~\frac{dv_k(N_k)}{dN}=-i \frac {k v_k}{a H_I}\Big|_{N=N_k}
\end{split}
\end{equation}
where $N_k\ll \log(k/(a H_I))$ is the horizon crossing e-folding and $a=k_*/H_I e^N$.

The power spectrum of curvature perturbations is then given by,
\begin{equation}
\Delta(k)= \frac {k^3}{2\pi^2} \left| \frac {v_k}{\sqrt{2 \epsilon}M_p a}\right|_{N=N_{\rm end}}^2
\end{equation}
Note that in the general case the quantities cannot be computed at horizon crossing but the evolution outside the horizon must be taken into account.
As we will verify in the SR regime the exact power spectrum is similar to the input one if $\eta\lesssim 1$ but this ceases to be true if the second SR condition is violated even if $\epsilon\ll 1$. Even in this case however the spectrum exhibits the required enhancement. Therefore, while a power spectrum cannot be in general mapped to a potential, for practical purposes of obtaining an enhanced power spectrum, the procedure above is sufficient even when SR conditions are violated. 

In what follows we  apply this method to the production of PBH and to the production of a stochastic GW background. 

\section{Production of PBH in slow roll and ultra slow roll}

As a first application we consider the production of PBH from inflationary fluctuations. This requires an enhanced scalar power spectrum compared to the CMB so that $\mathcal{P}(k_p)\sim 10^{-2}-10^{-3}$. The comoving momentum $k_p$ of the peak determines the typical mass of PBH produced,
\begin{equation}
M_{\rm PBH} \approx  \left(\frac{5.5 \times 10^{13}\, {\rm Mpc}^{-1}}{k_p}\right)^2 \times 10^{-15}\, M_{\odot}\,.
\label{eq:Mpbh}
\end{equation}
If PBHs constitute the totality of DM their mass is constrained in the asteroidal range $10^{-16}-10^{-12}\,M_\odot$, corresponding to perturbations that exited the horizon $O(35)$ e-foldings after the beginning of inflation. 

A popular way to realize the enhanced power spectrum is consider single field inflationary scenarios where the inflaton evolves through a very flat region, i.e. $V_{\phi}\approx 0$, see \cite{Ozsoy:2023ryl} for a review. Such scenarios are known as Ultra-Slow-Roll (USR) models. During USR $\epsilon \propto 1/a^6$ so that starting from the CMB amplitude $\sim 10^{-9}$, an enhanced power spectrum is quickly reached in just few e-foldings. For $V_\phi=0$ the transition regions violates the slow conditions as $\eta\approx 3$. 

Differently from USR we consider a variable growth of the power spectrum. 
We parametrize the input power spectrum as,
\begin{equation}
\mathcal{P}(k)= \mathcal{P}_{\rm CMB}(k)+\Delta \mathcal{ P}_{\rm CMB}(k)\,,~~~~
\Delta \mathcal{P}(k_*e^N)= \frac A 2\left[\tanh \frac{N-N_1}{w_1}+\tanh \frac{N_2-N}{w_2} \right]
\label{eq:parametrization}
\end{equation}
where we take $\mathcal{P}_{\rm CMB}(k)= A_s (k/k_*)^{-0.035}$ with $A_s=2.1\times 10^{-9}$ \cite{PL18}.
Before  the peak the spectrum is approximated by,
\begin{equation}
\Delta \mathcal{ P}\approx A \exp\left[2\frac{N-N_1}{w_1}\right]\approx A \left(\frac k {k_p}\right)^{2/w_1} 
\label{eq:IRtail}
\end{equation}
From eq. (\ref{eq:eta}) one can easily derive the second slow parameter $\eta\approx 1/w_1$. Therefore $w_{1,2}< 1 $ imply rapid transitions while for larger values the transition is smooth. In particular SR corresponds to $w_{1,2} \gtrsim 1$.  A growth comparable to USR is instead obtained for $w_1\sim 1/3$. Note that, differently from USR models, for $N\sim N_1$ the power spectrum depends logarithmically on $k$.

The spectrum deviates from $\mathcal{P}_{\rm CMB}$ for,
\be
N \gtrsim N_1+\frac {w_1} 2 \log \frac {\mathcal{P}_{\rm CMB}(k_*)}{A} \sim N_1- 7 w_1
\label{eq:deltaN}
\ee
For $w_1 \lesssim 1$ choosing $N_2\gtrsim N_1\gtrsim 10$ the amplitude is not modified at the CMB scale  but it approaches a maximum value $A$ for the modes that exit the horizon between $N_1$ and $N_2$ e-foldings of visible inflation. The transition regions are controlled by the widths $w_{1,2}$, corresponding to $\Delta N_{1,2} \sim 2 w_{1,2}$.

While other parametrizations are possible with our choice $\eta$ is roughly constant in the transition region. This allows to minimize the number of e-foldings necessary to build an enhanced power spectrum that will be relevant to build SR scenarios.

We now present two scenarios that produce the critical abundance of DM in the USR and SR regimes.
We choose parameters as in Table \ref{tab:asteroidal}.
$N_{1,2}$ are such that PBHs in the asteroidal mass range are produced. The amplitude $A$ is then chosen in order to reproduce the DM critical abundance after recomputing the exact power spectrum.

\begin{table}[h!]
\centering
\begin{tabular}{|c|c|c|c|c|c|}
\hline
\textbf{Model} & $N_1$ & $w_1$ & $N_2$ & $w_2$ & $A$ \\
\hline
\textbf{FAST} & 34.25 & 0.5 & 35.25 & 0.55 & $5.654\times 10^{-3}$ \\
\textbf{SLOW}  & 33.5 & 2 & 35.5 & 2 &$1.447\times 10^{-2}$\\
\hline
\end{tabular}
\caption{Parameters of the input power spectrum used to reproduce PBH abundance in USR and SR regimes.}
\label{tab:asteroidal}
\end{table}

\subsection{Fast Transition (PBH)}

\begin{figure}[h]
        \centering
    \begin{subfigure}{0.535\textwidth}
    \includegraphics[width=\textwidth]{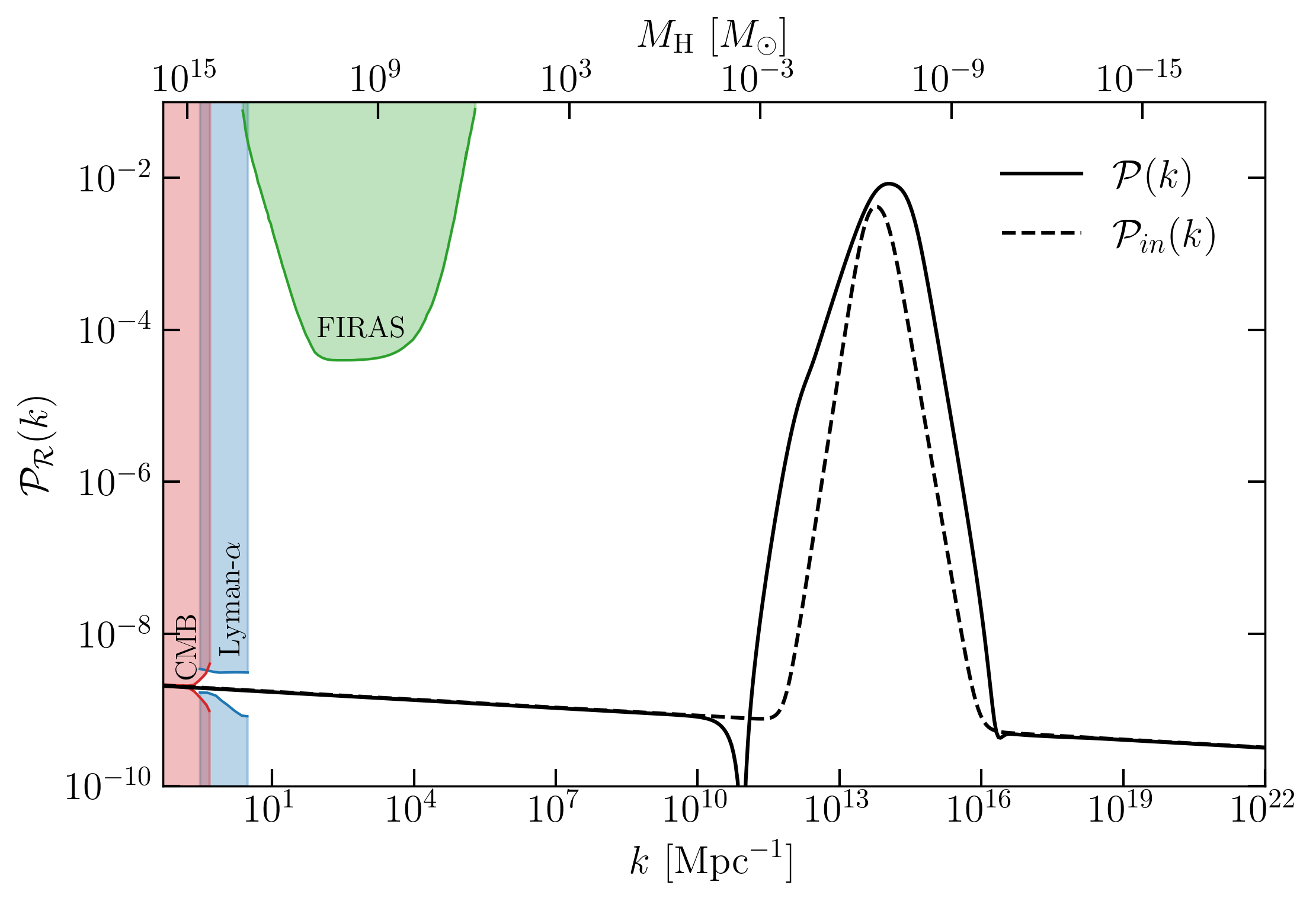} 
  \end{subfigure}
  \hfill
  \begin{subfigure}{0.45\textwidth}
    \includegraphics[width=\textwidth]{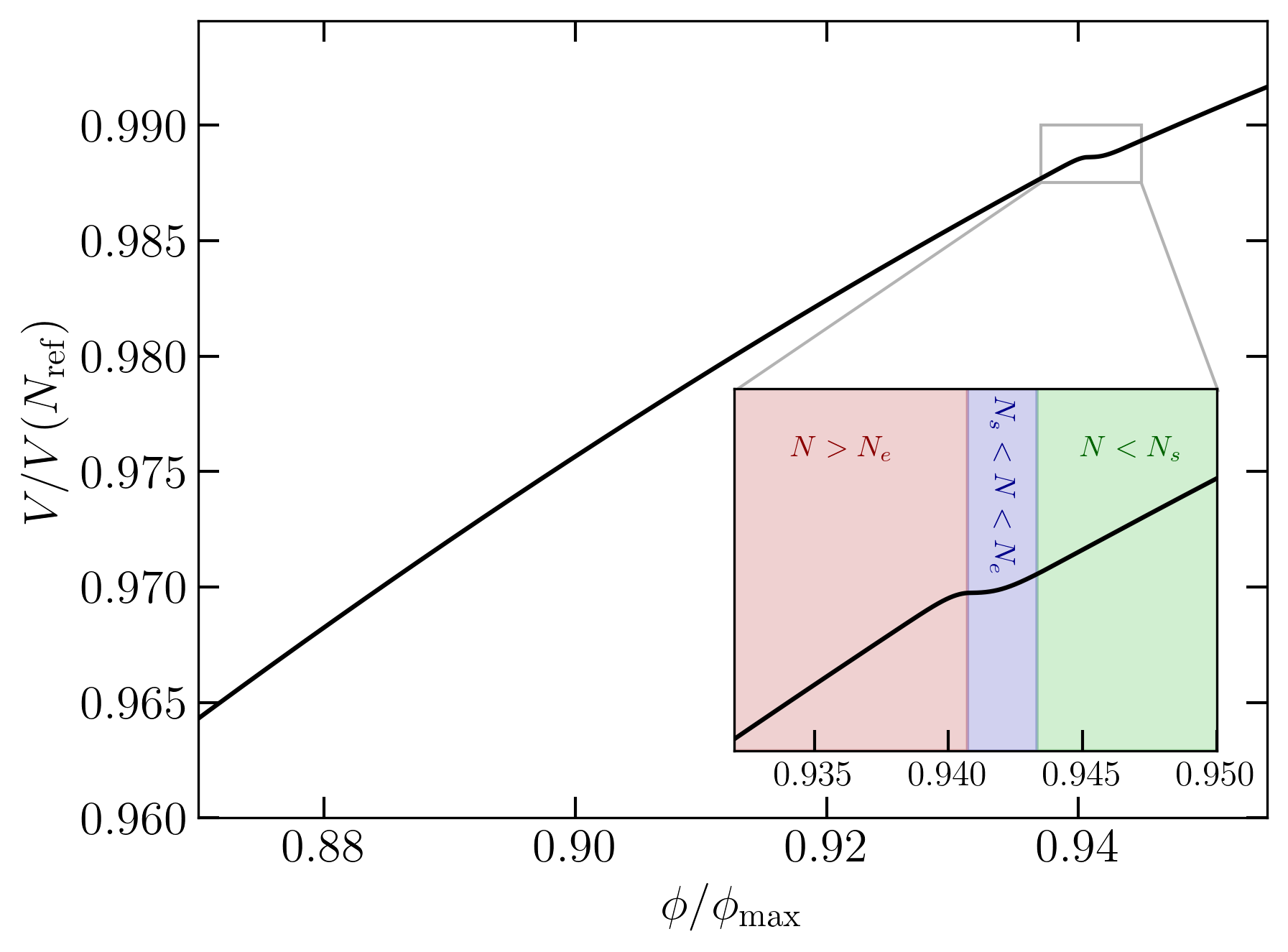}
  \end{subfigure}
    \caption{\textit{\textbf{Left Panel}}: Input power spectrum (dashed line) and the exact power spectrum for the fast transition model with parameters $N_1=34.25$, $N_2=35.25$, $w_1=0.5$, $w_2=0.55$, $A=5.654 \times 10^{-3}$. We plot the region excluded by measurements of CMB anisotropies \cite{PL18}, the FIRAS bound on CMB spectral distortions \cite{CE12} and the bound obtained from data on the Lyman-$\alpha$ forest \cite{BP11}. \textit{\textbf{Right Panel}}: Reconstructed potential.}  
    \label{PBH_fast_ps_pot}
\end{figure}

Using the parameters in table \ref{tab:asteroidal} for the fast transition model, the input power spectrum is shown with a dashed line in the left panel of Fig. \ref{PBH_fast_ps_pot}.  We reconstruct the potential solving numerically eqs. \eqref{eq:potential reconstruction}.  The resulting potential, as a function of the inflaton field $\phi$ is plotted on the right panel of Fig. \ref{PBH_fast_ps_pot}. On the right panel we show the parameters $\epsilon$ and $\eta$ computed with eqs.(\ref{eq:eps}) and (\ref{eq:eta}).\\
As expected, since $w_{1,2}<1$, the potential violates slow conditions in the transition regions where $|\eta|>1$. As a consequence one needs to recompute the power spectrum of perturbations. Using the exact values for $\epsilon(N)$ and $\eta(N)$ we  solve numerically eq. (\ref{eq:MS}) and derive the power spectrum  at the end of inflation. The result of the computation is shown in Fig.  \ref{PBH_fast_ps_pot}.
As expected the spectrum in the transition region is modified compared to the initial input because SR is violated. In particular prior to the enhancement the spectrum develops a dip that is a generic feature of USR scenarios. Nevertheless the presence of an enhancement follows directly from the input power spectrum.\\

\begin{figure}[h]
        \centering
    \begin{subfigure}{0.37\textwidth}
    \includegraphics[width=\textwidth]{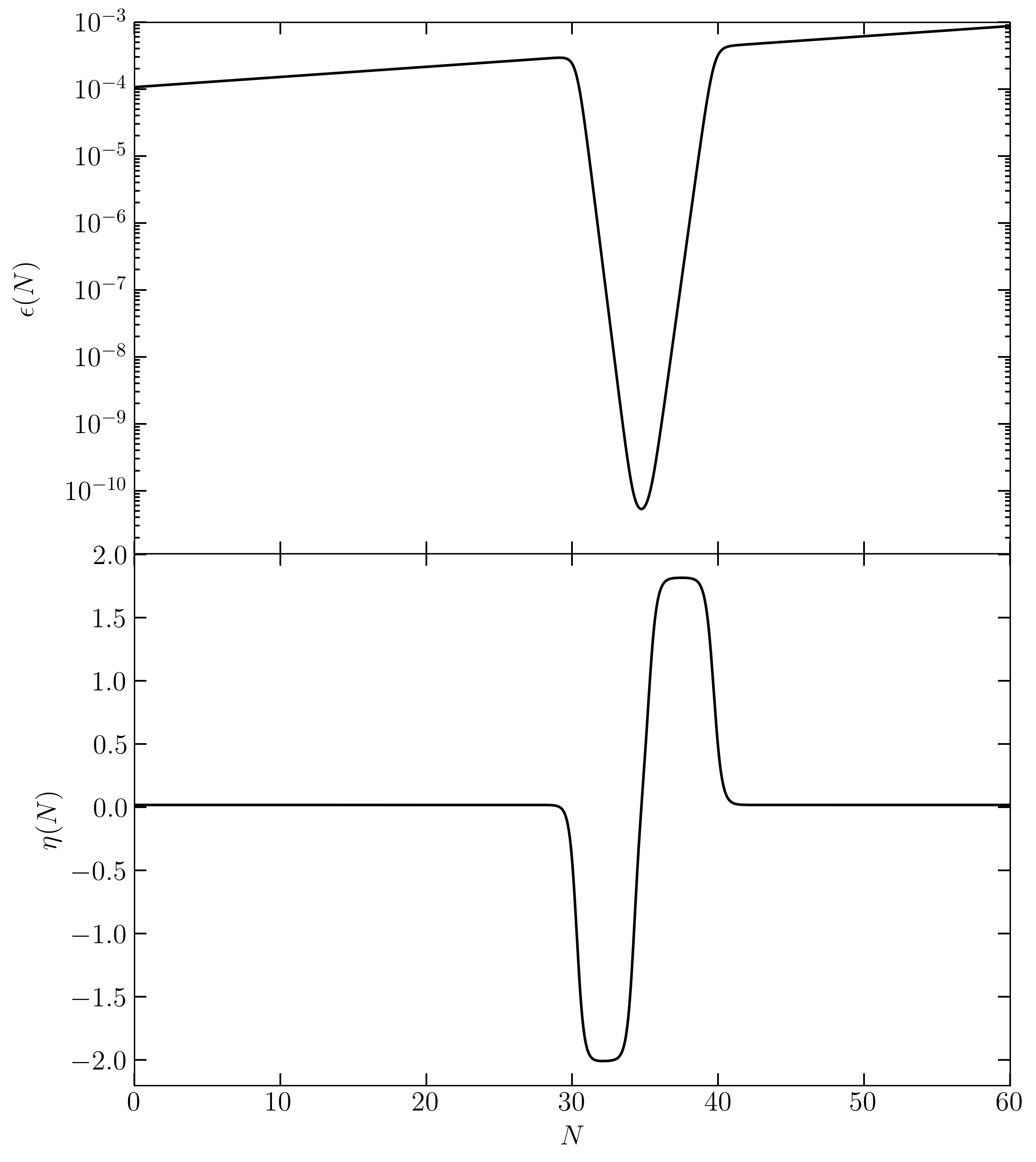} 
  \end{subfigure}
  \hfill
  \begin{subfigure}{0.62\textwidth}
    \includegraphics[width=\textwidth]{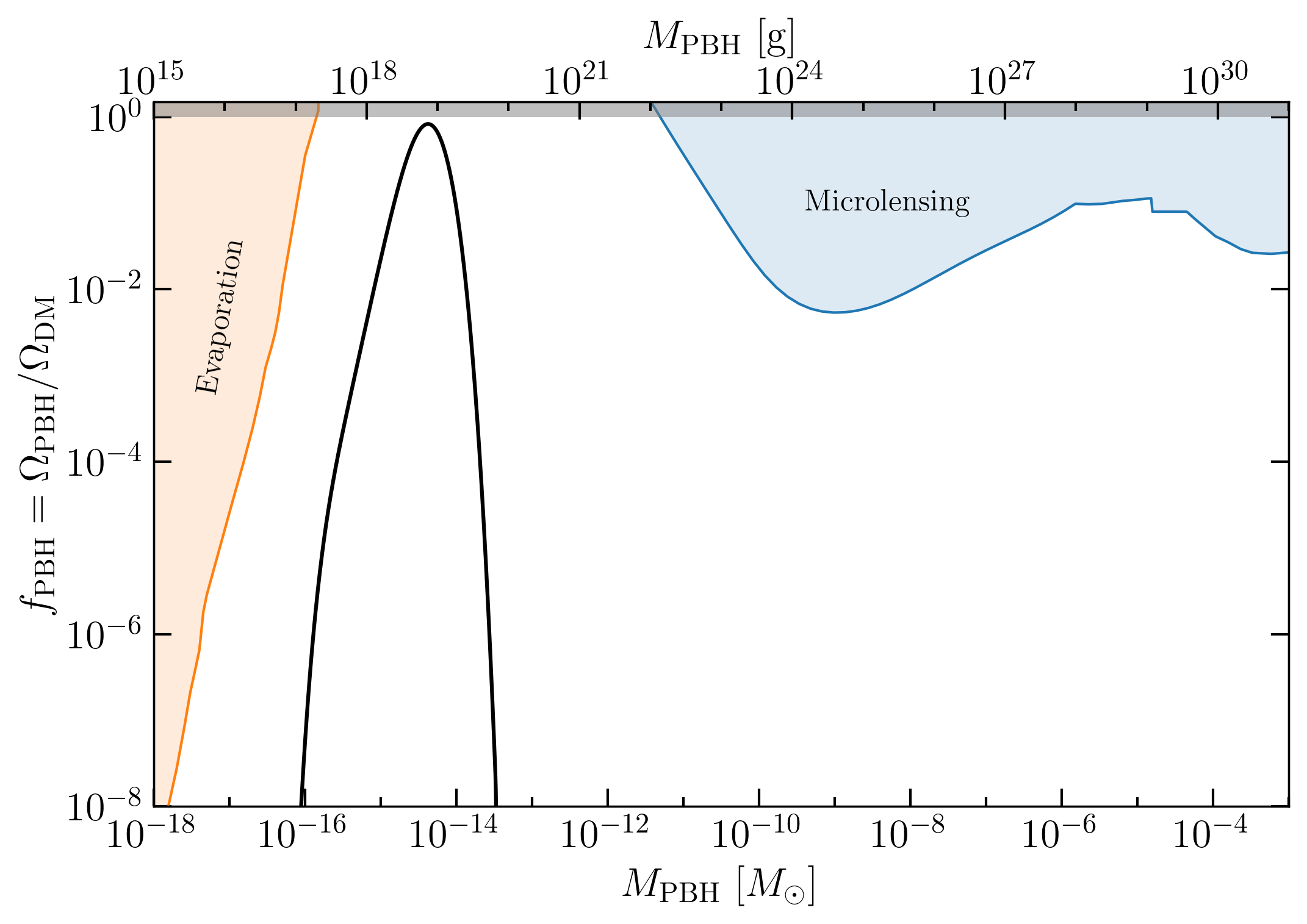}
  \end{subfigure}
    \caption{\textit{\textbf{Left Panel}}: Evolution of slow roll parameters $\epsilon(N)$ (\textit{\textbf{Top}}) and $\eta(N)$ (\textit{\textbf{Bottom}}).\\ \textit{\textbf{Right Panel}}: DM fraction in PBHs of mass $M_{\text{PBH}}$. For an updated repository of PBH bounds, see 
    \href{https://github.com/bradkav/PBHbounds}{/faGithub/bradkav/PBHbounds}. 
    In orange we plot evaporation bounds, in blue we plot micro-lensing bounds.}  
    \label{PBH_fast_epseta_pbh}
\end{figure}

With the exact power spectrum we then determine the abundance of PBH shown in the right panel of Fig. \ref{PBH_fast_epseta_pbh}. We follow here the procedure described in \cite{YM19,UF22,Frosina:2023nxu} to determine the critical threshold to form PBH in radiation domination. To compute the abundance we use \cite{UF22}
\begin{equation}
    f_{\text{PBH}}(M_{\text{PBH}}) = \frac{1}{\Omega_{\text{CDM}}}\int_{M_H^{\text{min}}}^{\infty}\frac{dM_H}{M_H}\biggl(\frac{M_{\text{eq}}}{M_H}\biggr)^{1/2}\biggl(\frac{M_{\text{PBH}}}{\mathcal{K}M_H}\biggr)^{1/\gamma}\biggl(\frac{M_{\text{PBH}}}{\gamma\,M_H}\biggr)\frac{\exp{-\frac{8[1-\sqrt{\Lambda}]^2}{9\sigma^2(M_H)}}}{\sqrt{2\pi\sigma^2(M_H)\Lambda}},
    \label{f_PBH}
\end{equation}
where
\begin{equation}
    \sigma^2(M_{H}) = \frac{16}{81}\int_0^{\infty}\frac{dk}{k} (kr_m)^4\,W^2(k,r_m) T^2(k,r_m)\mathcal{P}(k),
    \label{sigma_eq}\,,~~~    
\end{equation}
and $r_m(M_H)$ is the comoving length scale at which the so-called \textit{compaction function} has its maximum. Whether a perturbation collapses to form a PBH or not is ultimately determined by the amplitude of the perturbation at the maximum or the compaction function, that is, at scales $\approx r_m$. Furthermore, in equation \eqref{sigma_eq} we had that
\begin{equation}
  W(k,\tau) =  3\biggl[\frac{\sin{(k\tau)}-(k\tau)\cos{(k\tau)}}{(k\tau)^3}\biggr]\,~~~~~  \Lambda = 1-\frac{3\delta_c}{2}-\frac{3}{2}\biggl(\frac{M_{\text{PBH}}}{\mathcal{K}M_H}\biggr)^{1/\gamma}.
\end{equation}
The important threshold parameter is found to be $\delta_c=0.54$ for all the models explored, and we use $\mathcal{K} = 3.3$ and $\gamma = 0.36$.

\subsection{Slow transition (PBH)} 

\begin{figure}[h]
        \centering
    \begin{subfigure}{0.49\textwidth}
    \includegraphics[width=\textwidth]{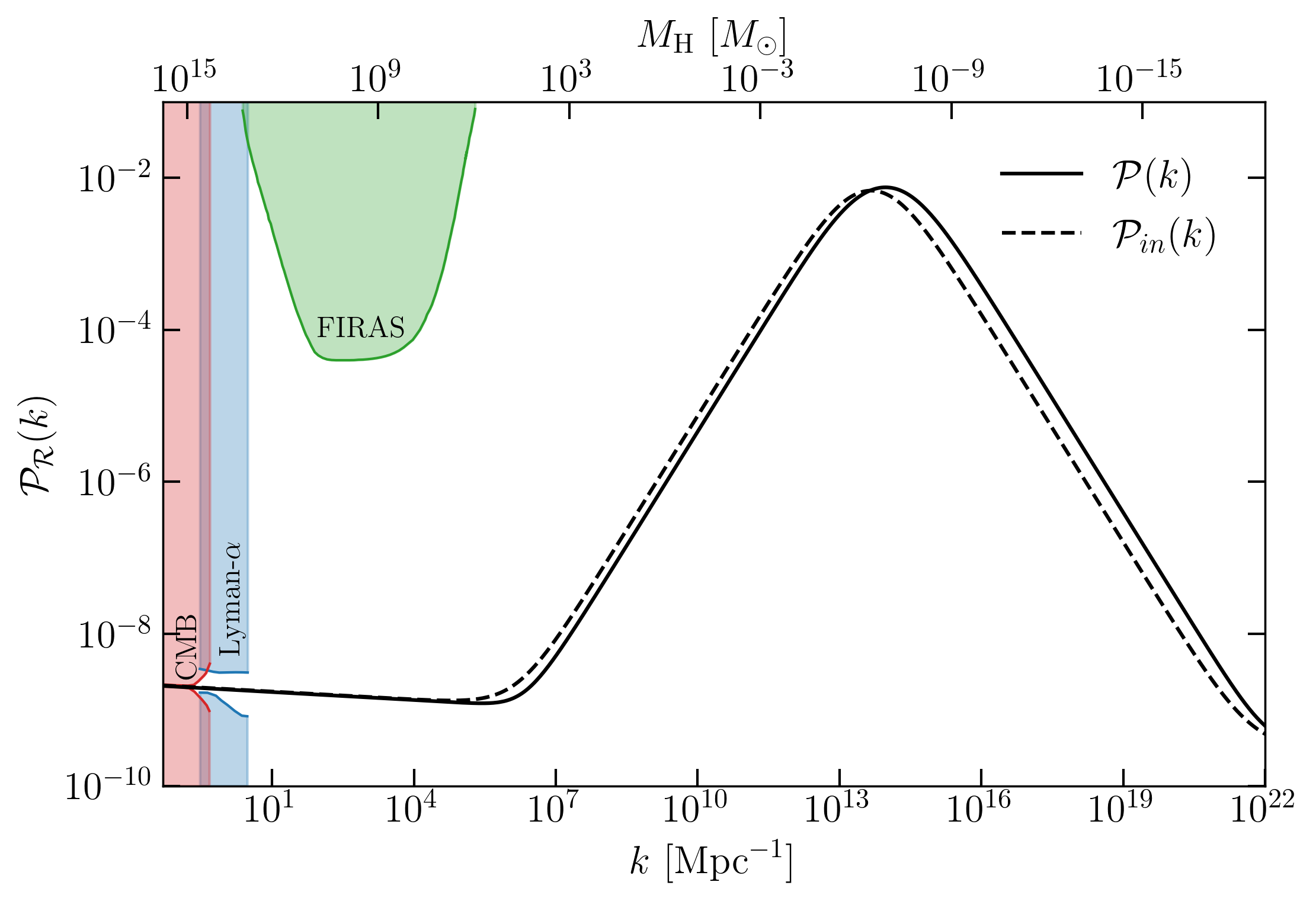} 
  \end{subfigure}
  \hfill
  \begin{subfigure}{0.44\textwidth}
    \includegraphics[width=\textwidth]{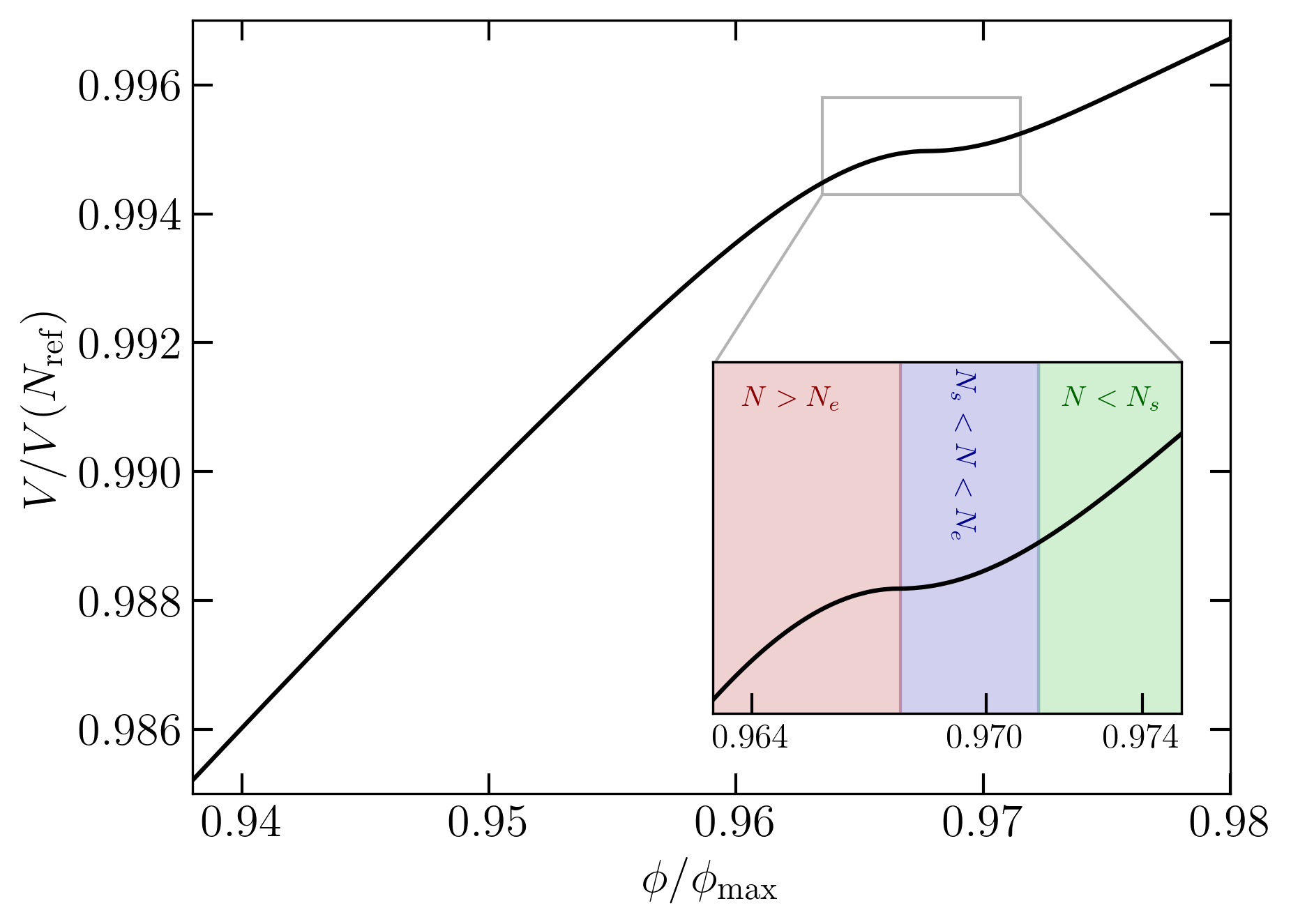}
  \end{subfigure}
    \caption{\textit{\textbf{Left Panel}}: Input power spectrum (dashed line) and the exact power spectrum for the slow transition model with parameters $N_1=33.5$, $N_2=35.5$, $w_1=2$, $w_2=2$, $A=1.447 \times 10^{-2}$. \textit{\textbf{Right Panel:}} Reconstructed potential.}
    \label{PBH_smooth_ps_pot}
\end{figure}

Next we consider a slow transition between the power spectrum at the CMB scales and the scale relevant for the production of PBH. Given the amplitude of the primordial power spectrum at the CMB scale, $\mathcal{P}(k_*)=2.1 \times 10^{-9}$ to produce significant fraction of black holes  the power spectrum must grow by almost 7 orders of magnitude. While $\epsilon$ can always be small from eq. (\ref{eq:eta}) the enhancement requires,
\begin{equation}
|\eta^{\rm max}| \Delta N> \frac 1 2 \log \frac {\epsilon_i}{\epsilon_f} \sim 7
\end{equation}
This estimate implies that the growth of the power spectrum requires $O(10)$ e-foldings if the transition takes place in the SR regime, $w_{1,2}\gtrsim1$.
Since asteroidal mass PBH are produced by perturbations that exited the horizon O(35) e-foldings after the beginning of inflation it follows that it should be possible to design potentials where the abundance of PBH is produced in the SR regime. \\
As a concrete example we consider the slow transition model with parameters give in Table \ref{tab:asteroidal}.
Using $\epsilon$ and $\eta$, we compute the exact power spectrum which we then compare to the input one in Fig. \ref{PBH_smooth_ps_pot} (solid black line). Even though $\eta\approx 0.5$ almost saturates the SR limit, the exact power spectrum is very similar to the input one, since the slow-roll approximation is never broken during inflation in this model. \\
In particular the dip feature that was present in the shark peak model is absent here, which again is due to the fact that the slow-roll approximation is not broken. This shows, contrary to \cite{Motohashi:2017kbs} that production of PBH is marginally compatible with SR inflation.

\begin{figure}[h]
        \centering
    \begin{subfigure}{0.37\textwidth}
    \includegraphics[width=\textwidth]{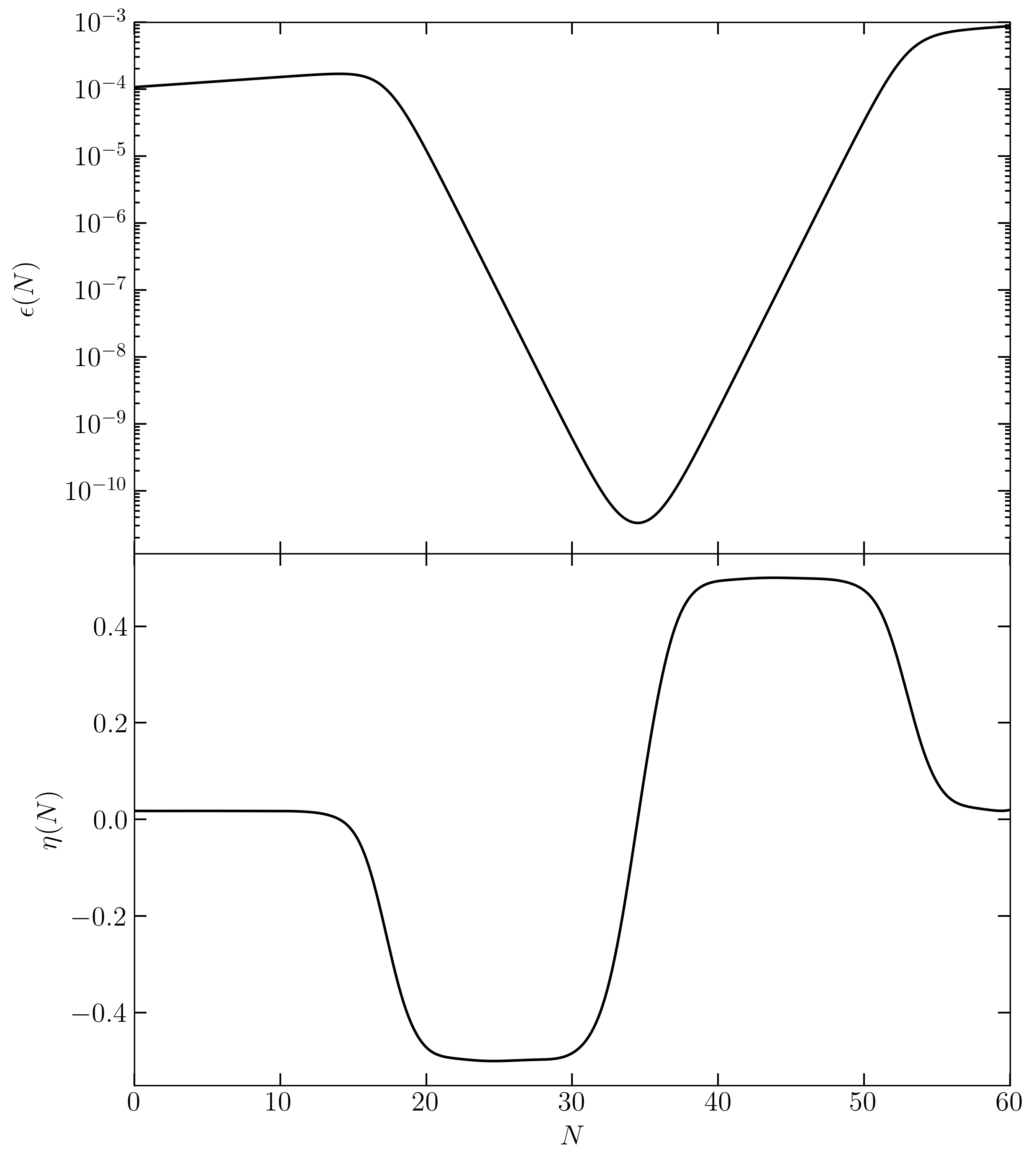} 
  \end{subfigure}
  \hfill
  \begin{subfigure}{0.62\textwidth}
    \includegraphics[width=\textwidth]{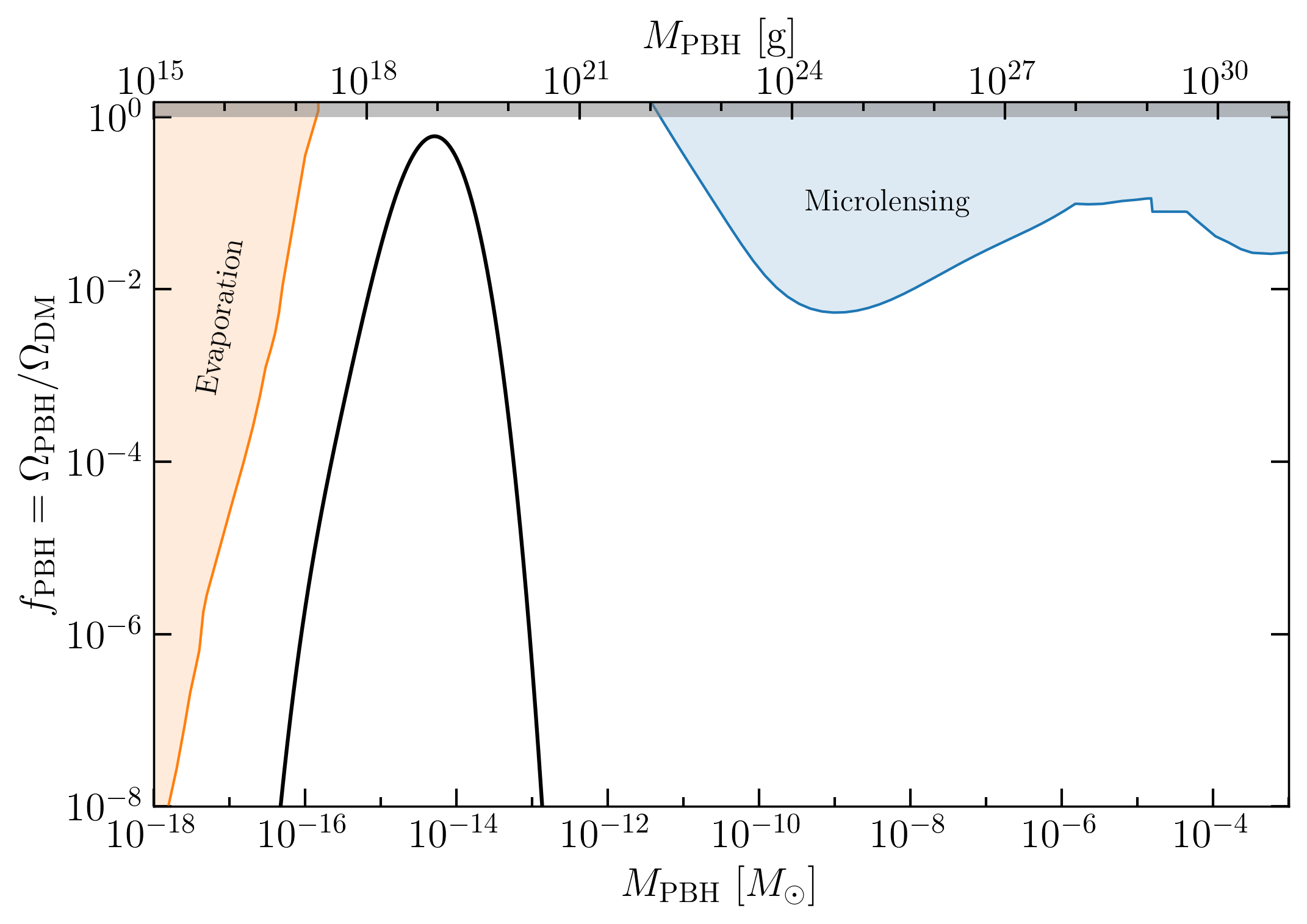}
  \end{subfigure}
    \caption{\textit{\textbf{Left Panel}}: Evolution of slow roll parameters $\epsilon(N)$ (\textit{\textbf{Top}}) and $\eta(N)$ (\textit{\textbf{Bottom}}).\\ \textit{\textbf{Right Panel}}: DM fraction in PBHs of mass $M_{\text{PBH}}$.}  
    \label{PBH_smooth_epseta_pbh}
\end{figure}

\vspace{18pt}
With the exact power spectrum, we compute the abundance of PBHs, which we show in Fig. \ref{PBH_smooth_epseta_pbh}. Despite the wider power spectrum the PBHs mass remains peaked similarly to the previous scenario. The amplitude of the power spectrum is negligible on cosmological scales so no significant constraints exist. For even larger $w_1$ larger modifications are found that could be testable in particular by $\mu-$distortions \cite{CE12} that are expected to improve in future experiments such as PIXIE \cite{Kogut:2011xw,Tagliazucchi:2023dai}.
\vspace{18pt}

\section{Gravitational waves at Pulsar Timing Arrays}

\label{sec:PTA}

An enhanced scalar power spectrum has also been proposed as the origin of the gravitational wave background recently observed in Pulsar Timing Arrays experiments \cite{NG15-1,NG15-2,NG15-3,NG15-4,EPTA-1,EPTA-2,EPTA-3,PPTA-1,PPTA-2,PPTA-3,CPTA}. While in single field inflation GWs produced by inflationary fluctuations are too small to explain the observed signal,  a larger amplitude can be generated as a secondary effect of large scalar perturbations \cite{Domenech:2020ers,MP94,AB03,MH04,AC07,BS07,Frosina:2023nxu,Domenech:2024rks,Choudhury:2024dzw,Heydari:2023rmq}. The same mechanism could also give observable signals at future experiments such as LISA \cite{LISA} if the scalar power spectrum is sizable in the mHz region.

In light of the data a good fit requires a scalar power spectrum with a blue tilt. We here show how to construct such a model both in the USR and SR regime.

\subsection{Computation of the Gravitational Wave Background}

The stochastic GW background is fully determined from the scalar power spectrum \cite{Espinosa:2018eve}. We follow Ref. \cite{DO21} for the computation of the abundance of GWs, and refer to that for details. The abundance of GWs today is given by
\begin{equation}
    \Omega_{\text{GW}}(f) = \frac{c_g \Omega_r}{36}\int_0^{\frac{1}{\sqrt{3}}} dt\int_{\frac{1}{\sqrt{3}}}^{\infty} ds\,\mathcal{T}(t,s)\,\mathcal{P}\biggl[\frac{k\sqrt{3}}{2}(s+t)\biggr]\mathcal{P}\biggl[\frac{k\sqrt{3}}{2}(s-t)\biggr],
    \label{gw_abund}
\end{equation}
where 
\begin{equation}
    \mathcal{T}(t,s)\equiv \biggl[\frac{(t^2-1/3)(s^2-1/3)}{t^2-s^2}\biggr]^2\bigl[\mathcal{I}_c^2(t,s)+\mathcal{I}_s^2(t,s)\bigr],
\end{equation}
Here $\Omega_r$ is the energy density of radiation today, $\mathcal{I}_c$ and $\mathcal{I}_s$ are analytic functions
\begin{equation}
\begin{split}
\mathcal{I}_c(t,s)&=-36\pi\frac{\left(s^2+t^2-2\right)^2}{\left(s^2-t^2\right)^3}\,\theta(s-1),\\
\mathcal{I}_s(t,s)&=-36\frac{\left(s^2+t^2-2\right)}{\left(s^2-t^2\right)^2}\left[\frac{\left(s^2+t^2-2\right)}{\left(s^2-t^2\right)}\log{\frac{(1-t^2)}{\vert s^2-1\vert}}+2\right].
\end{split}
\end{equation}
and
\begin{equation}
    c_g\equiv \frac{g_{\ast}(T)}{g_{\ast}^0}\biggl[\frac{g_{\ast S}^0}{g_{\ast S}(T)}\biggr]^{4/3}.
    \label{c_g}
\end{equation}
accounts for the change of the effective degrees of freedom with temperature $T$.  The temperature dependence of $g_{\ast}$ and $g_{\ast S}$ can be found for example in \cite{SK18}. 
It is well known that deep in radiation domination, $c_g \approx 0.4$, however modes that are relevant for the NANOGrav signal re-enter the horizon close to the QCD phase transition, resulting in a value of $c_g$ which is slightly larger than the radiation value. Moreover the modified equation of state can have an important impact on the GWs
abundance leading to an enhancement at frequencies relevant for PTAs \cite{Abe:2020sqb,Franciolini:2023wjm,Domenech:2024rks}.

The frequency of the GW can be related to the 
temperature where the modes re-enter the horizon as,
\begin{equation}
    f \sim  10^{-8} \biggl(\frac{T}{\text{GeV}}\biggr)\,\text{Hz},
    \label{f_of_T}
\end{equation}
so that the background is expeted to be produced at a temperature of order GeV. This corresponds to modes that
exited the horizon around,
\begin{equation}
N \sim 19 + \log \frac {f}{10^{-8}\,{\rm Hz}}\,.
\label{eq:ftoN}
\end{equation}

Two features should be noted.
Obviously rescaling the amplitude of the scalar perturbations by a factor $A$ changes the GW power spectrum by a factor $A^2$. From the formula above it also follows that for $\mathcal{P}(f/f_*)$ the GW background is  also function of $f/f_*$.  As a consequence one can translate the spectrum to different scales by simply shifting the scalar power spectrum. This observation will be useful to fit a power spectrum to the experimental signal.

%At the same time, a given mode $k$ re-enters the horizon at a temperature $T$ which is given by the relation
%\begin{equation}
%    k = 1.5\times 10^7 \biggl(\frac{g_{\ast}}{106.75}\biggr)^{1/6}\biggl(\frac{T}{\text{GeV}}\biggr)\,\text{Mpc}^{-1},
 %   \label{k_of_T}
%\end{equation}
%which allows us to trade the dependence on temperature in the expression for $c_g$ for a dependence on wave-number $k$. Furthermore, all dependencies on $k$ or $f$ in equation \eqref{gw_abund} can be traded for one another by means of the relation
%\begin{equation}
%    k \simeq 6.47\times 10^{14}\biggl(\frac{f}{\text{Hz}}\biggr)\,\text{Mpc}^{-1}.
%    \label{k_of_f}
%\end{equation}

\begin{table}[h]
\centering
\begin{tabular}{|c|c|c|c|c|c|}
\hline
\textbf{Model} & $N_1$ & $w_1$ & $N_2$ & $w_2$ & $A$ \\
\hline
\textbf{FAST} & 18 & 0.65 & 25.5 & 1.6 & $1.06\times 10^{-2}$\\
\textbf{SLOW}  & 18 & 1.8 & 24 & 2.5 & $1.08\times 10^{-2}$ \\
\hline
\end{tabular}
\caption{Parameters of input power spectrum employed to reproduce NANOGrav signal.}
\label{tab:GWparam}
\end{table}

In this work we focus  on NANOGrav \cite{NG15-1,NG15-2,NG15-3,NG15-4},but similar analysis can be carried out for other datasets. NANOGrav is sensitive to a GW background in the range $f\subset[2\cdot 10^{-9}\,, 6\cdot 10^{-8}]$ Hz. From eq. (\ref{eq:ftoN}) this corresponds to $N\subset[17\,,20.5] $. This is also the range where the scalar power spectrum should be enhanced to produce the signal.

We now present 2 scenarios that reproduce the signal with the parameters given in Table \ref{tab:GWparam}. These parameters were chosen to reproduce the NANOGrav signal and an abundance of PBH compatible with current constraints.

\subsection{Fast transition (GW)}

\begin{figure}[h]
        \centering
    \begin{subfigure}{0.49\textwidth}
    \includegraphics[width=\textwidth]{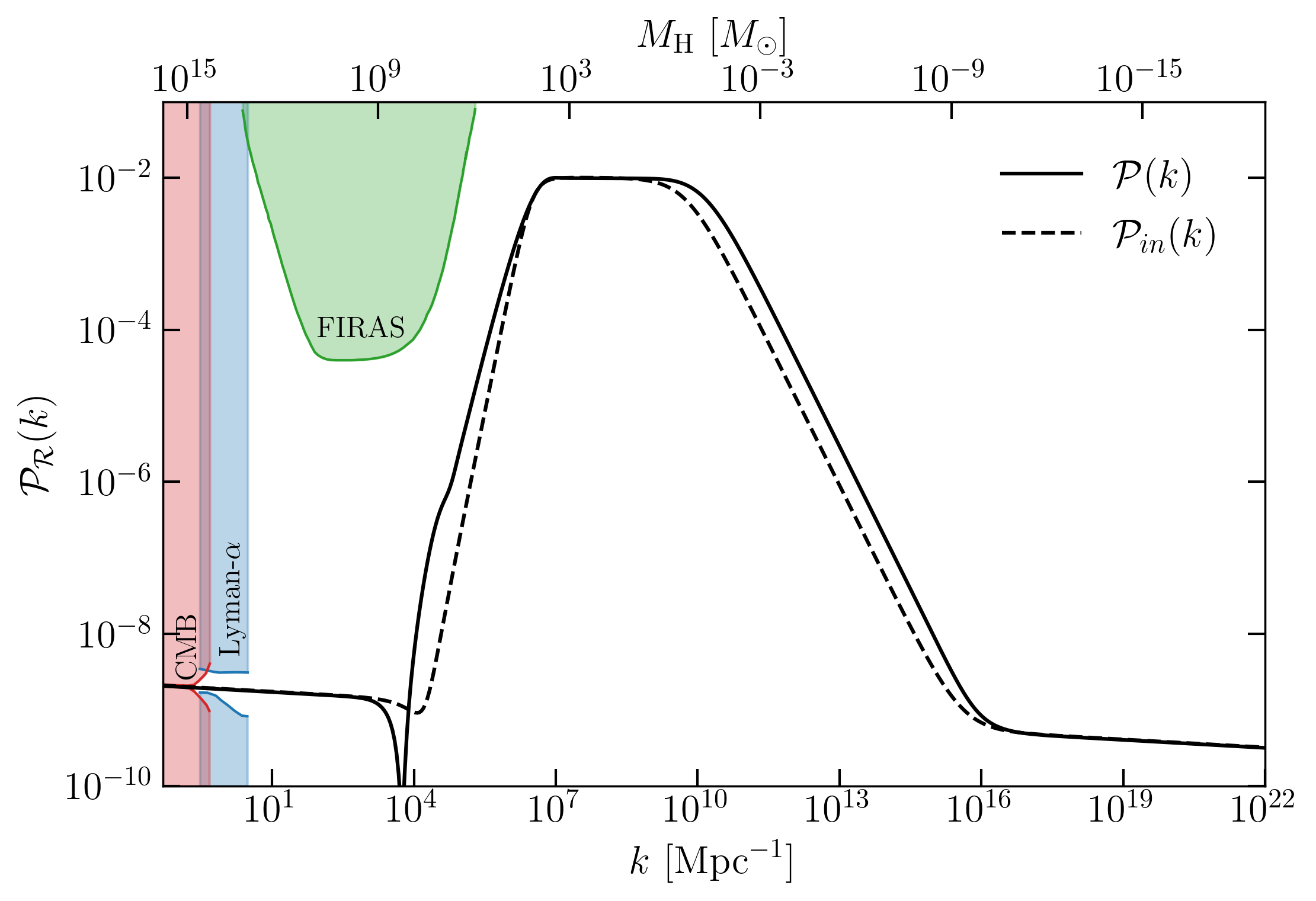} 
  \end{subfigure}
  \hfill
  \begin{subfigure}{0.44\textwidth}
    \includegraphics[width=\textwidth]{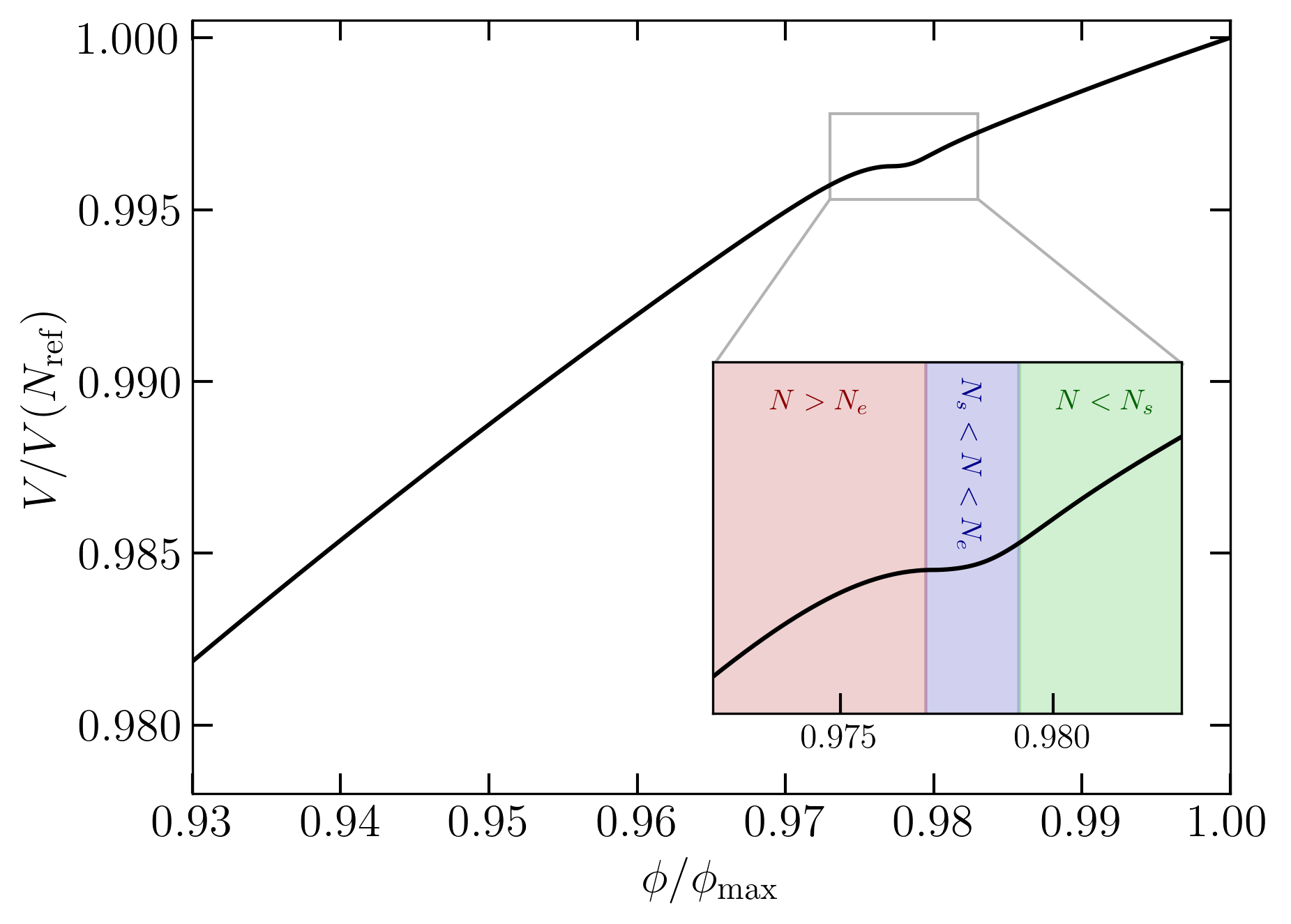}
  \end{subfigure}
    \caption{\textit{\textbf{Left Panel}}: Input power spectrum (dashed line) and the derived exact power spectrum that reproduce the NANOGrav signal for a fast transition model with with parameters $N_1=18$, $N_2=25.5$, $w_1=0.65$, $w_2=1.6$, $A=1.06 \times 10^{-2}$. \textit{\textbf{Right Panel}}: Reconstructed potential.}
    \label{GW_fast_ps_pot}
\end{figure}

We first consider the scenario with a rapid growth of power spectrum that deviates from SR in the transition region. Given that the GW signal must be blue tilted we are mostly interested in the raising part of the scalar power spectrum that is controlled by $N_1$ and $w_1$. Note in particular that the peak is reached at $N_1+w_1$ that should be of order 20 to generate GWs in the PTA range.

\begin{figure}[h]
\centering
    \begin{subfigure}{0.50\textwidth}
    \includegraphics[width=\textwidth]{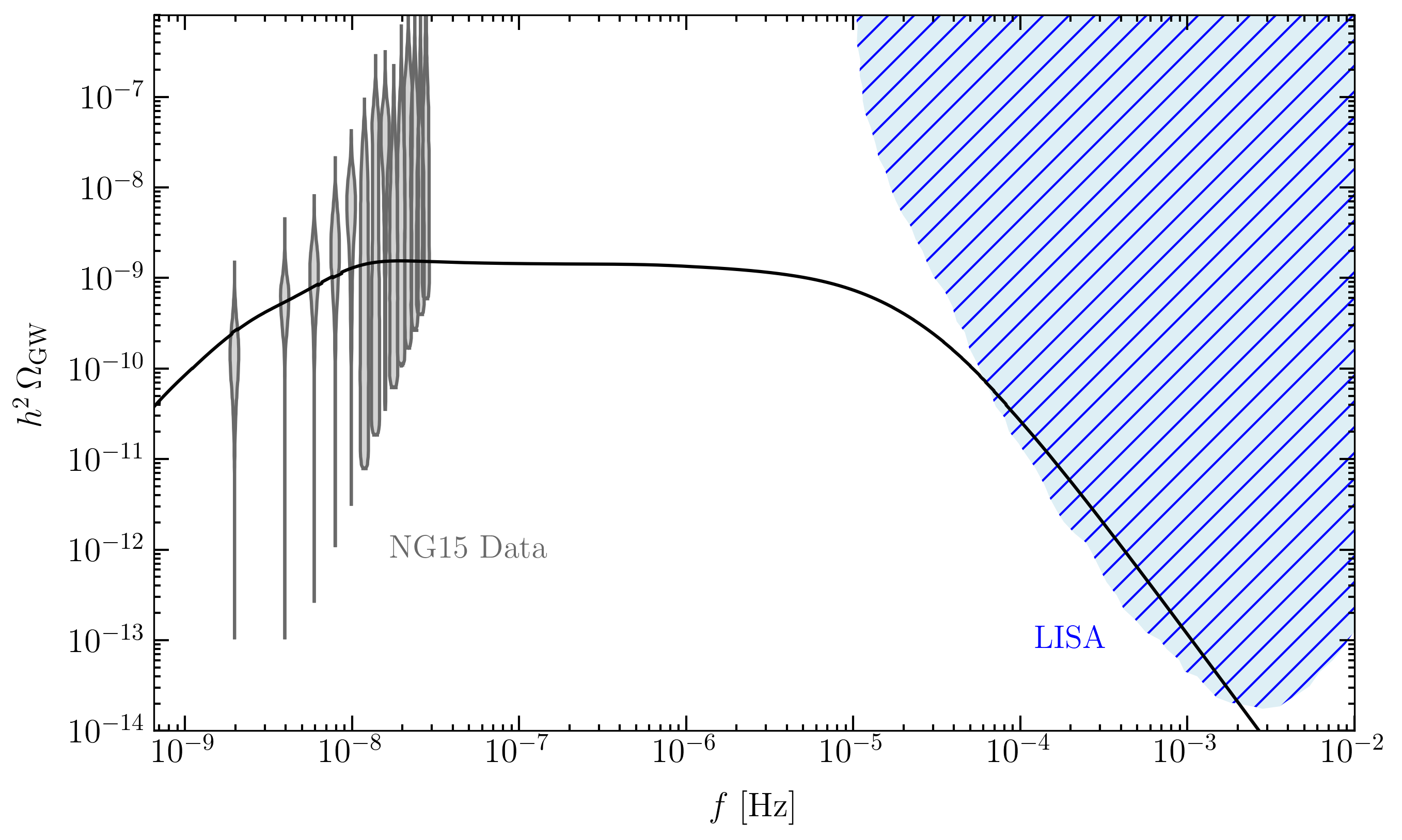} 
  \end{subfigure}
  \hfill
  \begin{subfigure}{0.475\textwidth}
    \includegraphics[width=\textwidth]{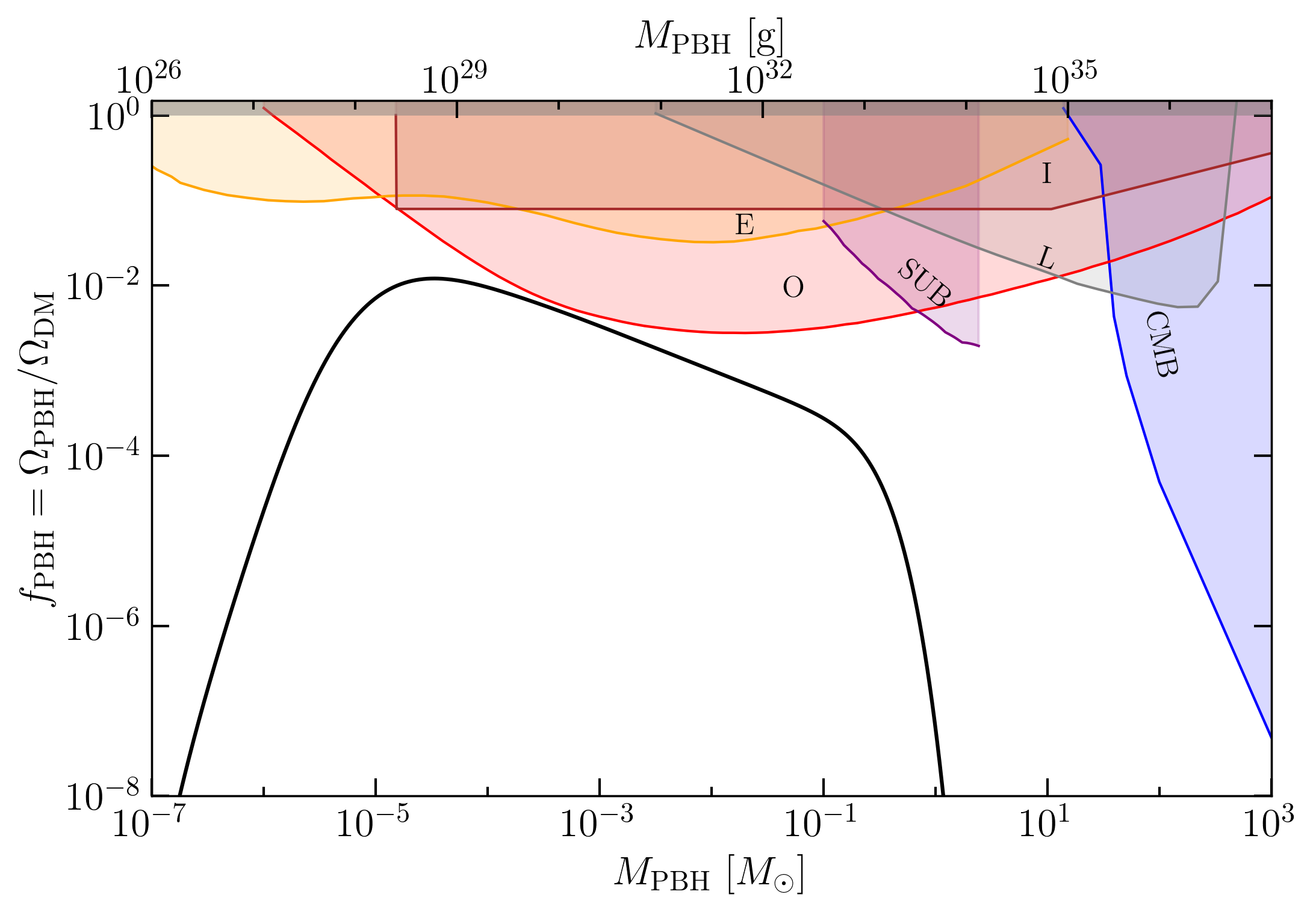}
  \end{subfigure}
\caption{\textit{\textbf{Left Panel}}: Fraction of energy density in gravitational waves for the fast transition model. The gray violins are the first 14 bins of the NANOGrav 15-year data set, the blue region is the power-law integrated sensitivity curve of future space-based GW interferometer LISA \cite{LISA}. \textit{\textbf{Right Panel}}: PBH DM fraction for the sharp transition model. We plot the following bounds: Microlensing constraints from: OGLE-III and OGLE-IV (\textbf{O}) \cite{OGLE1,OGLE2}, EROS (\textbf{E}) \cite{ER07}, Icarus (\textbf{I}) \cite{IC18}; constraints from GWs searches from coalescence of sub-solar mass binaries (\textbf{sub}) \cite{SUB21}; direct constraints from PBH-PBH mergers with LIGO (\textbf{L}) \cite{LI19,LI18}; constraints from modifications of CMB spectrum due to accreting PBHs (\textbf{CMB}) \cite{CA20}.}
    \label{GW_fast_gw_pbh}
\end{figure}

A good fit of the signal is obtained for $N_1=18$ and $w_1=0.65$ and weakly depends on $N_2$ and $w_2$. We choose $N_2$ significantly larger than $N_1$ to produce a wide plateau where the power spectrum is peaked. With these parameters, we apply our reconstruction procedure to compute the exact scalar power spectrum and from this we determine the induced secondary GW background. In Fig. \ref{GW_fast_ps_pot} we show the input scalar power spectrum and the derived one while the GW background is shown Fig. \ref{GW_fast_gw_pbh} and compared with data.\\
Let us note that for rapid growth of the scalar power spectrum the resulting GW background has an IR tail proportional to $k^3$ as implied by causality while the data prefer a milder slope \cite{NG15-1}. However if the signal is fitted in the region close to the maximum of the potential a good fit can anyway be found. \\
The signal in fact must be fit not very far from the maximum for a different reason. The enhanced scalar power spectrum implies that also PBHs will be produced. The frequency scales relevant for NG15 correspond to PBHs in the sub-solar mass range, where strong bounds on the fraction of DM exist. 
As shown in Fig. \ref{GW_fast_gw_pbh} for the parameters chosen the abundance of PBH saturates current bounds. Increasing the amplitude of the scalar power spectrum would thus result in a tension between the mass function of PBHs and the microlensing bound placed by the EROS collaboration \cite{ER07} and the bound placed by GWs searches due to the coalescence of sub-solar mass PBH binaries \cite{SUB21}. We also note that, at the same time, the position of the peak plays a crucial role. Indeed, moving the peak of the scalar power spectrum towards larger scales (smaller $k$) would result in a $f_{\text{PBH}}(M_{\text{PBH}})$ which would be peaked at bigger mass scales, resulting in a tension with the SUBARU bound.\\
In this model the parameters have been chosen so that the GW spectrum is large over a wide range of frequencies as in Ref. \cite{DeLuca:2020agl}. As shown in Fig. \ref{GW_fast_gw_pbh} the signal is within the sensitivity of future GW experiments such us \cite{LISA}. 
\vspace{18pt}

\subsection{Slow Transition (GW)}

\begin{figure}[h]
        \centering
    \begin{subfigure}{0.49\textwidth}
    \includegraphics[width=\textwidth]{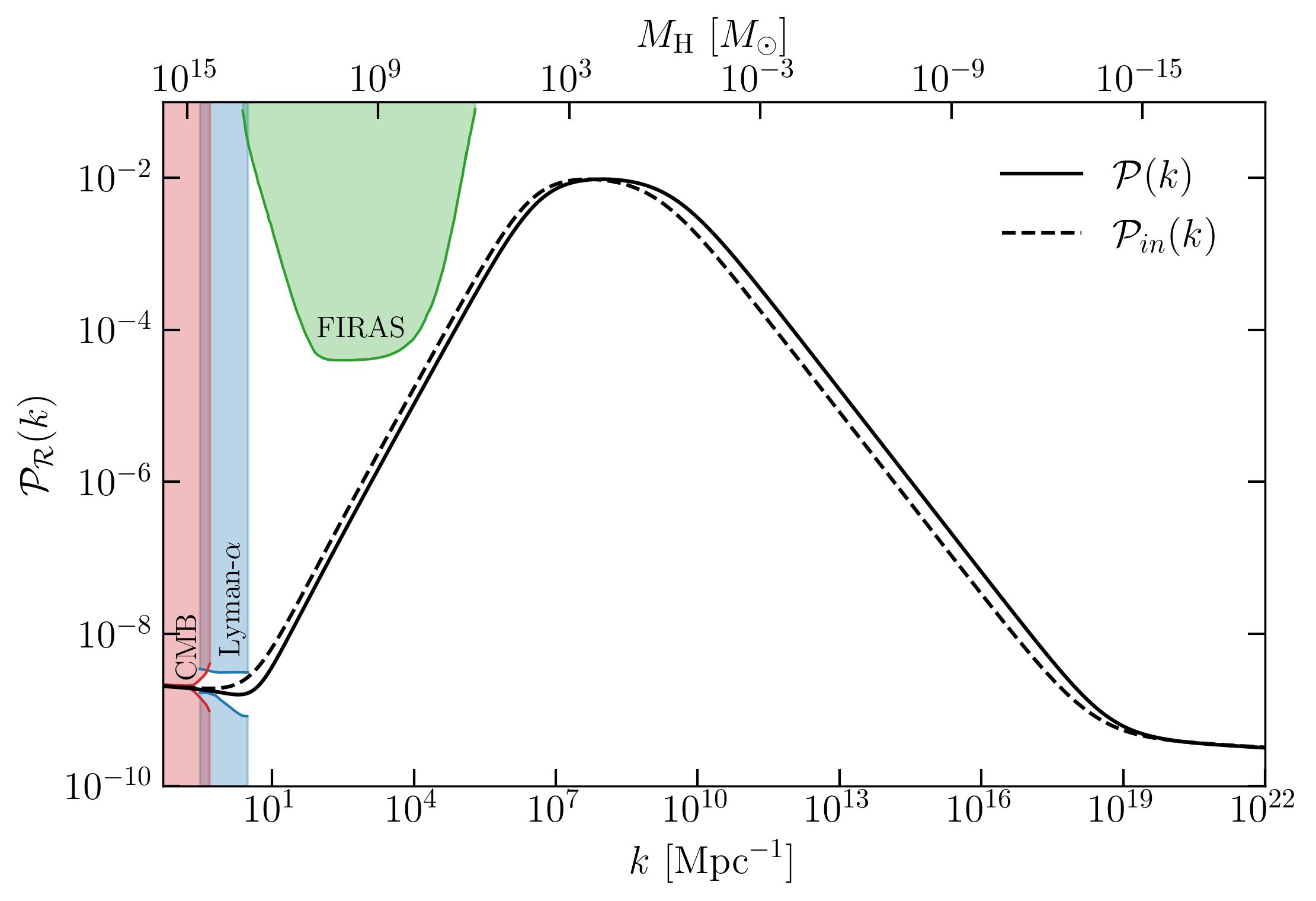} 
  \end{subfigure}
  \hfill
  \begin{subfigure}{0.44\textwidth}
    \includegraphics[width=\textwidth]{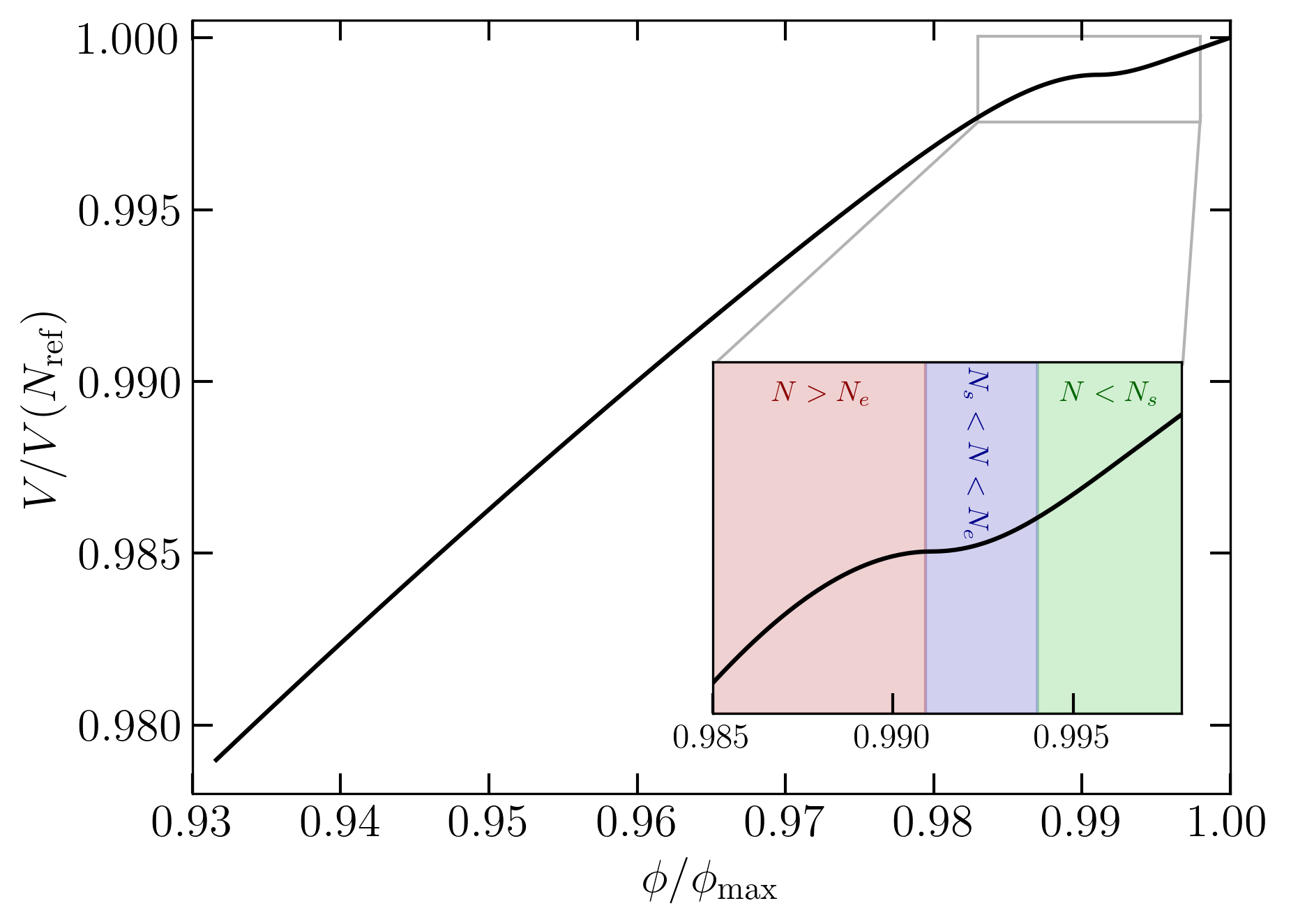}
  \end{subfigure}
    \caption{\textit{\textbf{Left Panel}}: Input scalar power spectrum (dashed line) and the exact power spectrum for a slow transition model with parameters $N_1=18$, $N_2=24$, $w_1=1.8$, $w_2=2.5$, $A=1.08 \times 10^{-2}$. \textit{\textbf{Right Panel}}: Reconstructed potential.}
    \label{GW_smooth_ps_pot}
\end{figure}
\vspace{18pt}
Let us now consider the possibility that the transition to the enhanced power spectrum takes place in the SR regime. 
For the parameters in Table \ref{tab:GWparam} the input and  exact power spectrum of the reconstructed potential are plotted in Fig. \ref{GW_smooth_ps_pot}. 

In this case the power spectrum  grows within the SR approximation and peaks at scales relevant for NANOGrav, while barely avoiding current bounds from $\mu-$distortions as well as the constraints the Lyman-$\alpha$ forest. The IR tail of the scalar power spectrum (\ref{eq:IRtail}) is  particularly relevant for the PTA signal where the spectrum must be enhanced at much larger scales than for asteroidal PBH. The scalar power spectrum is compatible with the CMB for $w_1 \lesssim 3$.  A stronger bound arise from $\mu-$distortions that test the power spectrum at shorter scales even though less precisely. Future measurements from PIXIE \cite{Kogut:2011xw,Tagliazucchi:2023dai} will improve the constraints by orders of magnitude and will thus be able to test this type of scenario. 
Recently even stronger constraints have been proposed by Graham and Ramani \cite{Graham:2024hah} from heating of stars. At face value their bound would currently exclude a large fraction of parameter space of model. We leave a detailed analysis to future work.

\begin{figure}[h]
\centering
    \begin{subfigure}{0.50\textwidth}
    \includegraphics[width=\textwidth]{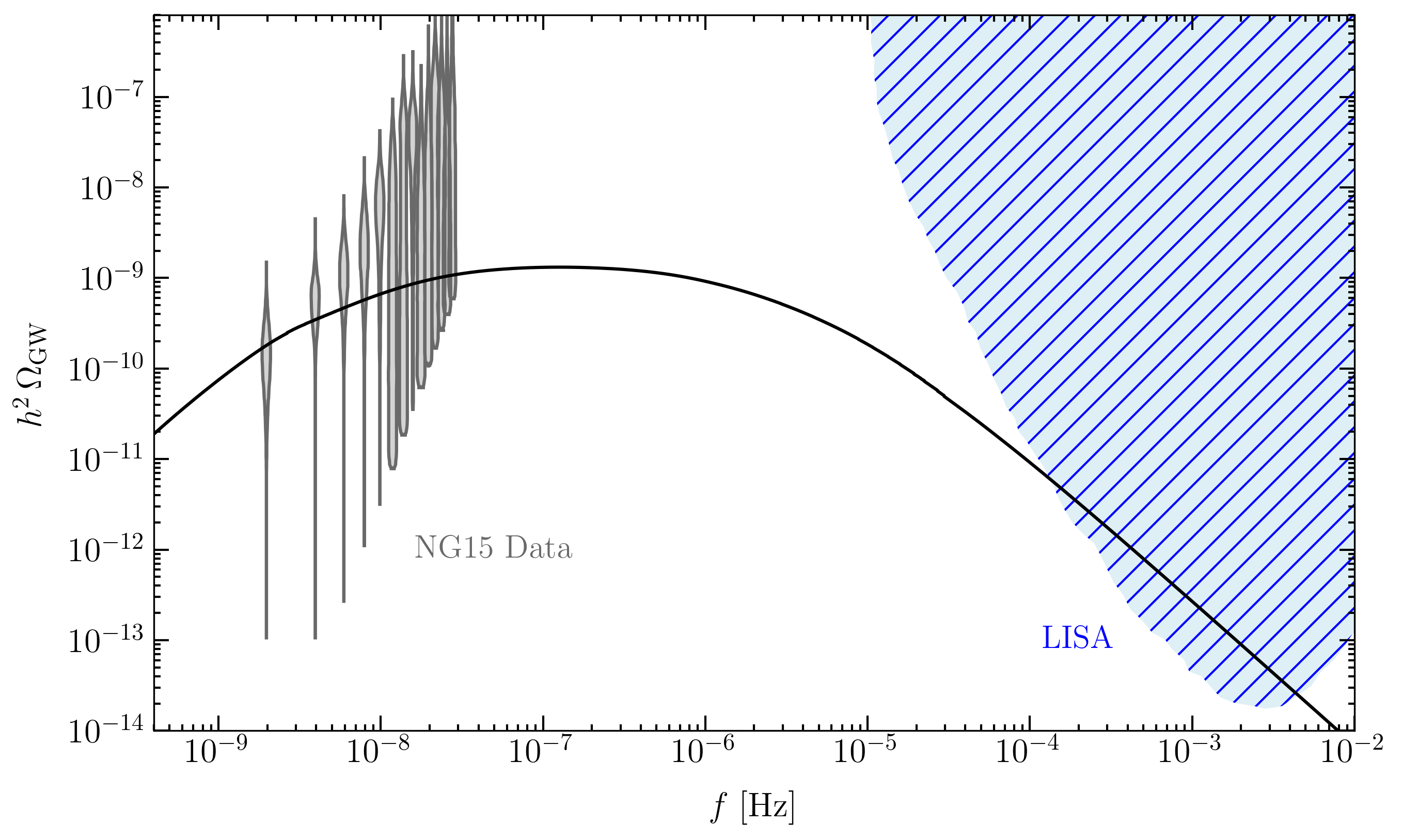} 
  \end{subfigure}
  \hfill
  \begin{subfigure}{0.475\textwidth}
    \includegraphics[width=\textwidth]{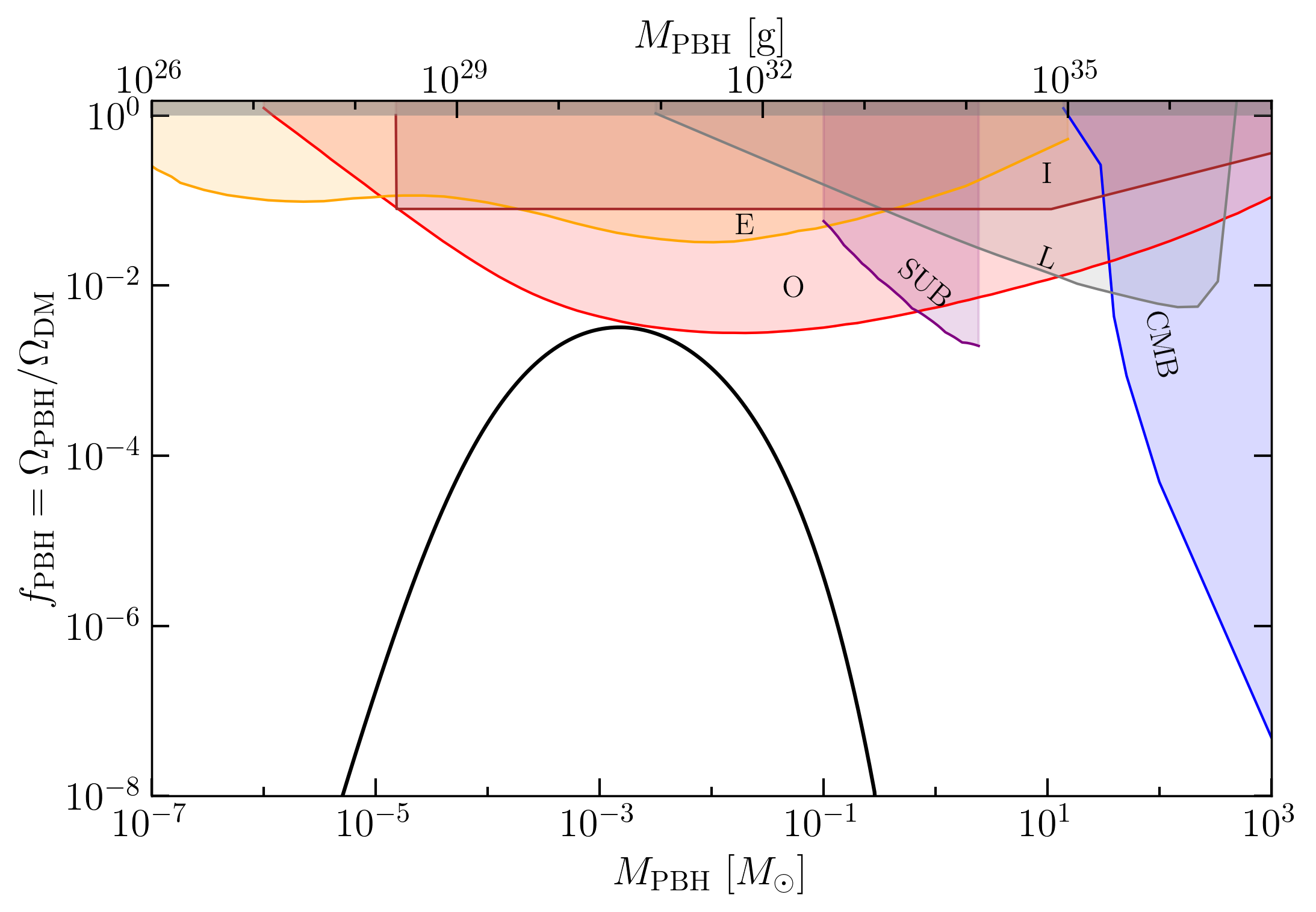}
  \end{subfigure}
\caption{\textit{\textbf{Left Panel}}: Energy fraction in gravitational waves for the smooth transition model. The grey violins are the NG15 data points, in blue we report the LISA power-law integrated sensitivity curve. \textit{\textbf{Right Panel}}: DM fraction in PBHs of mass $M_{\text{PBH}}$ for the  model with smooth transition. We plot the same bounds as in Fig. \ref{GW_fast_gw_pbh}.}
    \label{GW_smooth_gw_pbh}
\end{figure}

In Fig. \ref{GW_smooth_gw_pbh} we show the abundance of GW produced. We note a broader profile compared to the model with fast transition. 
In the case of slow transition the IR tail of the GW power spectrum is just the square of the scalar power spectrum. Indeed $\Omega_{\rm GW}\sim k^{4/w_1}$ in the IR for $w_1> 4/3$. A good approximation  of the power spectrum is given by
\begin{equation}
\Omega_{\rm GW}\approx  C \left[\tanh \frac{N-N_1}{w_1}+\tanh \frac{N_2-N}{w_2} \right]^2
\label{parametrization}
\end{equation}
where $C\sim 10^{-4} A^2$. 

\begin{figure}[h]
    \centering
    \includegraphics[width=0.6\linewidth]{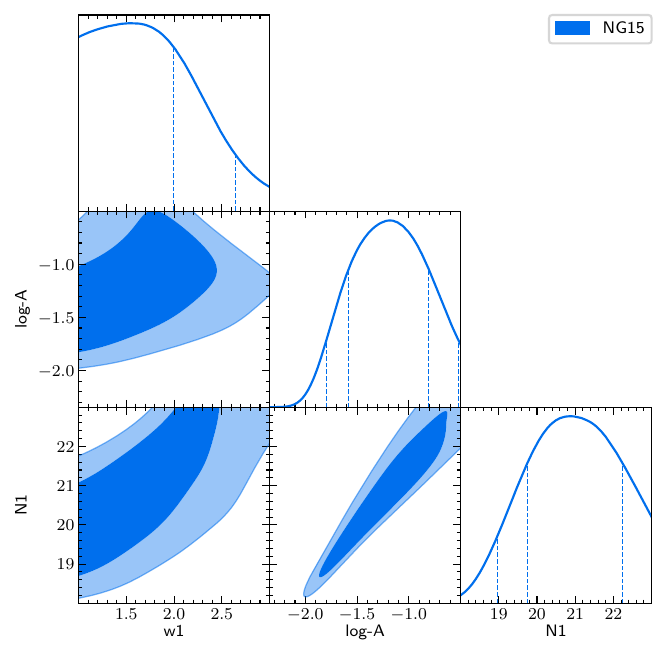}
    \caption{Marginalized posterior distributions for the model parameters derived with \texttt{PTArcade}  using NANOGrav 15 years dataset. We use as prior $w_1>1$ where the slow roll approximation is justified.}
    \label{fig:ptarcade}
\end{figure}

To conclude we consider a Bayesian analysis of the model  using the code \texttt{PTArcade} \cite{Mitridate:2023oar,lamb2023need}. Assuming the parametrization (\ref{eq:parametrization}) we can find the allowed regions of parameters that reproduce the NANOGrav signal. Given that the signal is only mildly sensitive to the $w_2$ with fix it to $w_2=2.5$ and take $N_2-N_1=6$. For simplicity we do not recompute the power spectrum with the procedure used in the rest of the paper as in the SR regime it is very similar. The posterior distributions of parameters are shown in Fig. \ref{fig:ptarcade}. In particular we see that NANOGrav favours $A> 10^{-2}$ (best fit is found for $w_1=1.3$, $N_1=20$ and $A=0.04$).  
On the other hand the model considered in this section has $A\approx 10^{-2}$ that as shown in the right panel of Fig. \ref{GW_smooth_gw_pbh} saturates the constraints on the abundance of solar mass PBHs. This confirms the tension between reproducing the nominal value of the amplitude suggested by  NANOGrav and the abundance of solar mass PBHs \cite{Dandoy:2023jot,Franciolini:2023pbf,Andres-Carcasona:2024wqk,Iovino:2024uxp}.

\section{Summary and outlook}

In this note we presented a general method to construct single field inflationary models with desired features focusing on an enhanced power spectrum.  The idea is to start with the desired power spectrum and reconstruct the potential from that. To define the model we determine $\epsilon$ in the SR regime and reverse engineer the potential that produces it. This procedure would be exact in the SR regime but the method remains effective when the SR conditions are violated such as in USR models. In this case the exact power spectrum is different from the input one in the transition region but continues to reproduce the enhancement. 

We applied this method to construct models with an enhanced power spectrum both in the SR and USR regime. Such a power spectrum is interesting because it leads to the production of PBH that could reproduce the abundance of DM in our universe. We showed in  particular that this is possible in the SR regime while most studies in the past have focused on USR models. The enhanced scalar power spectrum has also been advocated to reproduce the signal observed in Pulsar Timing Arrays experiments that recently detected a stochastic GW background at  nano-Hertz frequencies. This requires a power spectrum enhanced at $k\sim 10^5\,{\rm Mpc}^{-1}$ that would be produced after $N\sim 20$ e-foldings of visible inflation. In this case the tail of the power spectrum can lead to observable effects in the CMB and $\mu$ distortions that allow to place strong constraints in the SR regime. 

Recently the consistency of inflationary scenarios with an enhanced power spectrum has been questioned starting with Ref. \cite{Kristiano:2022maq}. The enhanced power spectrum at short scales could induce at 1-loop a correction to the power spectrum at large scales potentially in tension with CMB measurements of the amplitude and tilt. A clear consensus on the size of the correction and whether it can be reabsorbed in the parameters of the model has not yet been reached even though at least in some cases the correction appears to be small. The discussion has been mostly within the framework of USR models where analytic solutions for the wave-functions are known. In the case of a slow transition the calculation should be reconsidered because the contribution is not localized at a given time. Our preliminary study shows that the result can be different in this case but it would require a better understanding on the UV dependent part of the contribution. We hope to return to this question in future work. 

\subsubsection*{Acknowledgements}
MR would like to thank Andrea Mitridate for support with PTArcade. GA is supported by INDARK INFN PD51 grant.

\pagestyle{plain}
\bibliographystyle{jhep}
\small
\bibliography{biblio}

\end{document}